\documentstyle[12pt]{article}

\global\arraycolsep=1pt
\begin{document}
\def\o#1#2{{#1\over#2}}
\def\bfone{\relax{\rm 1\kern-.35em 1}}
\def\inbar{\vrule height1.5ex width.4pt depth0pt}
\def\IC{\relax\,\hbox{$\inbar\kern-.3em{\rm C}$}}
\def\ID{\relax{\rm I\kern-.18em D}}
\def\IF{\relax{\rm I\kern-.18em F}}
\def\IH{\relax{\rm I\kern-.18em H}}
\def\II{\relax{\rm I\kern-.17em I}}
\def\IN{\relax{\rm I\kern-.18em N}}
\def\IP{\relax{\rm I\kern-.18em P}}
\def\IQ{\relax\,\hbox{$\inbar\kern-.3em{\rm Q}$}}
\def\IR{\relax{\rm I\kern-.18em R}}
\font\cmss=cmss10 \font\cmsss=cmss10 at 7pt
\def\ZZ{\relax\ifmmode\mathchoice
{\hbox{\cmss Z\kern-.4em Z}}{\hbox{\cmss Z\kern-.4em Z}}
{\lower.9pt\hbox{\cmsss Z\kern-.4em Z}}
{\lower1.2pt\hbox{\cmsss Z\kern-.4em Z}}\else{\cmss Z\kern-.4em
Z}\fi}
\hfill    HUTP-94/A041
\par
\hfill    SISSA 182/94/EP
\par
\hfill    hepth@xxx/9411205
\par
\hfill   November, 1994
\par
\begin{center}
\vspace{10pt}
{\large \bf
GAUGED HYPERINSTANTONS\\
AND\\
MONOPOLE EQUATIONS\\}
\vspace{30pt}
{ Damiano Anselmi\,\footnotemark\footnotetext{Partially supported
by the Packard Foundation and by NSF grant PHY-92-18167.}}
\par
\vspace{4pt}
{\it Lyman Laboratory, Harvard University,
Cambridge MA 02138, U.S.A.}
\par
\vspace{20pt}
{ Pietro Fr\`e\,\footnotemark\footnotetext{Partially supported by EEC,
Science Project SC1$^*$-CT92-0789.}}
\par
\vspace{4pt}
{\it SISSA - ISAS, via Beirut 2-4,
34014 Trieste, Italia \\
INFN - Sezione di Trieste, Trieste, Italia}
\par
\vspace{30pt}

{\bf Abstract}
 \end{center}
The monopole equations in the dual abelian theory
of the N=2 gauge--theory, recently
proposed by Witten as a new tool to study topological
invariants, are shown to be the simplest elements in a class of
instanton equations that
follow from the improved topological twist mechanism
introduced by the authors
in previous papers.
When applied to the N=2 $\sigma$--model, this twisting procedure
suggested the introduction of the so-called hyperinstantons that are the
solutions to an appropriate condition of triholomorphicity imposed
on the maps $q : {\cal M} \, \rightarrow \, {\cal N}$ from a
four--dimensional almost quaternionic world--manifold ${\cal M}$ to an almost
quaternionic target manifold ${\cal N}$. When gauging the
$\sigma$--model by coupling it to the vector multiplet of a gauge
group $G$, one gets instantonic conditions (named by us {\sl
gauged hyperinstantons}) that reduce to the Seiberg--Witten equations
for ${\cal M}={\cal N}=\IR^4$ and $G=U(1)$.
The deformation of the self--duality
condition on the gauge--field strength due to the
monopole--hyperinstanton
is very similar to the deformation of the self--duality condition
on the Riemann curvature previously observed by the authors when
the hyperinstantons are coupled to topological gravity. In this
paper the general form of the hyperinstantonic equations coupled
to both gravity and gauge multiplets is presented.

\vspace{4pt}

\noindent
\eject
Recently
the non perturbative results obtained
by Seiberg and Witten
\cite{topf4d_2,topf4d_3}
on the infrared behavior of
N=2 gauge theories for a gauge group $G$
have attracted a lot of interest.
The N=2 supersymmetric pure Yang--Mills theory has
a moduli--space of vacua, namely it admits
flat directions of the
scalar potential, and there has been a concentration
of efforts on studying the geometry of this
space \cite{topf4d_2,topf4d_3,topf4d_5,topf4d_7,topf4d_6}.
This is done by considering the effective
lagrangian which, if N=2 supersymmetry is
preserved, must fall into the general form of an
N=2 super Yang--Mills lagrangian for the unbroken
gauge subgroup $H \, \subset \, G$. This is completely
encoded in the choice of a
{\it flat special K\"ahler geometry} structure
\cite{skgsugra_2,skgsugra_4,skgsugra_1,skgmat_1,topf4d_5,skgsugra_11},
namely into a holomorphic section
$\left \{ X^{i}(z) \, , \,
\o{\partial {\cal F}(X)}{\partial X^{i}}
\right \}$ of a flat $Sp({\rm dim}\, H,\IR)$
bundle, determining the kinetic K\"ahler
metric of the vector multiplet scalars $X^{i}$
via the formula
$g_{ij^\star}=\o{\partial}{\partial z^{i}} \,
\o{ \partial}{\partial {\bar z}^{i^\star}}
\left ( X^{i}\, \partial_{i^\star}
{\bar {\cal F}}(\bar X )
+{\bar X}^{i^\star}\, \partial_{i} { {\cal F}}(X)
\right )$. The non--perturbative determination
of the holomorphic section
$\left \{ X^{i}(z) \, , \,
\o{\partial{\cal F}(X)}{\partial X^{i}}
\right \}$ is performed by relying on duality
considerations that connect the infrared and the
ultraviolet regimes, by inverting the strength of
the gauge coupling constant and exchanging
magnetic with electric charges. The implementation
of a discrete group of duality transformations leads
to a set of Picard--Fuchs equations for
$\left \{ X^{i}(z) \, , \,
\o{\partial {\cal F}(X)}{\partial X^{i}}
\right \}$, that are indeed interpreted as {\it
periods} of suitable holomorphic forms on the
moduli space. This is just analogous to what happens in
Calabi--Yau
compactifications \cite{picfucperiod_2,skgpicfuc_2}.
\par
 One starts from a {\it
microscopic theory} that is a pure N=2
gauge--theory for the group $G$ with the choice,
for its flat special geometry,
of the
{\it minimal coupling}
${\cal F}(X)=\sum_{i=1}^{{\rm dim}\, G}\,( X^{i})^2$
and one arrives at an {\it effective dual theory}
which is also an N=2 gauge theory displaying,
however, the following differences.

i) The gauge
group is abelian and it is the dual
${\tilde H}$  of the maximal torus
$H\, \subset \, G$ in the original gauge group.

ii) The self--interaction of the gauge--multiplet
is encoded in a {\it non--minimal}
{\it flat special geometry} possessing a discrete
group of duality symmetries.

iii) In addition
to the gauge--multiplet the theory contains a
certain number of {\it N=2 hypermultiplets} that
represent the {\it monopoles} of the original
theory.

The last point in this list of properties
is the main issue and motivation of the present
letter, in conjunction with a recent
suggestion by Witten \cite{topf4d_4}.

It is well-known that there is a relation
between topological Yang-Mills theory
\cite{topfgen_7,topfgen_5} and the mathematical
problem of calculating Donaldson invariants
of four--manifolds \cite{topfgen_4,topfgen_8}. As a matter
of fact, this has longly been believed to be an
{\sl equivalence} relation, in the sense that Donaldson invariants
were thought to be the only physical amplitudes of
topological Yang-Mills theory. However, it has been
recently shown by one of us \cite{me} by explicit solving
the theory in the case $M=\IR^4$, $G=SU(2)$ and unit
instanton number, that certain anomalous behaviors are able to
enrich the theory with many nonvanishing amplitudes computing
{\sl link} invariants. In this sense, topological Yang-Mills theory
can no longer be considered equivalent to Donaldson theory.

Moreover, in \cite{me} computations of a third kind of
topological invariants of four manifolds were performed.
These are some (again anomalous) physical amplitudes
of topological gravity, eventually coupled
to topological Yang-Mills theory. These invariants were
constructed in \cite{topftwist_1,topftwist_2}
in a general context.
Finally, a fourth type of topological invariants
of  four manifolds are those related
to the topological $\sigma$-model, constructed in \cite{topf4d_1}.

On the other hand, topological
Yang Mills theory, just as topological gravity
\cite{topftwist_1} and the
topological sigma model
\cite{topftwist_2,topf4d_1}, can be obtained
by topologically twisting an N=2 gauge--theory.
Hence, due to the conjectured equivalence \cite{topf4d_2,topf4d_3}
between the infrared
limit of the N=2 gauge--theory of $G$ with its
{\it dual}  N=2 theory of the abelian group
${\tilde H}$ (coupled to monopoles), the suggestion of \cite{topf4d_4} is to
recast
the problem of calculating topological invariants of whatever type
into a {\it dual abelian framework}. To this purpose, it is worth
considering the {\it topological twist}
of the {\it dual theory}. Although Witten did not
phrase his argument exactly in these terms, we think
that this is more or less the point. Indeed,
the whole idea of \cite{topf4d_4} is that {\it ``rather than computing
the Donaldson invariants by counting
$SU(2)$ instanton solutions, one can obtain the
same invariants by counting the solutions
of the dual equations, which involve $U(1)$
gauge fields and monopoles"}.
As a matter of fact, we think that both the original
and the dual gauge theory, once topologically twisted,
provide a framework for the calculation of topological
invariants that are quite worth consideration. Their
relation is possibly still a matter of debate, but
certainly those associated with the dual theory are
of the utmost interest. It is therefore important,
in our opinion, to clarify the general meaning of
the monopole equations mentioned in the above
quotation from Witten's paper.

Effective lagrangians are not constrained
by power counting renormalizability.
On the other hand, topological field theories are finite
and exactly soluble \cite{dam,dam2,me}.
So, they represent a very general and powerful tool for
studying topological invariants of four manifolds
with the methods of physics \cite{me}.

What are these {\it monopole equations}? In \cite{topf4d_4}, using
considerations on spin--bundles and focusing
on the dual  of the minimal $SU(2)$ theory,
namely on an N=2 $U(1)$ gauge--theory coupled
to one hypermultiplet,
Witten obtained the following equations:
\begin{equation}
F_{\alpha\beta}={i\over 2}(M_\alpha \bar M_\beta+
M_\beta \bar M_\alpha),
\quad \quad D_{\alpha\beta^\prime}M^\alpha=0.
\label{mono}
\end{equation}
where $M_\alpha$, $\alpha=1,2$ are the two
complex scalars belonging to the {\it monopole
hypermultiplet}, while $F_{\alpha\beta}$ is the
antiself--dual part of the $U(1)$ gauge field
written in a formalism that uses spinor
indices.

In the present letter we want to show that:
\par
i) Eq.s (\ref{mono}) are just
the {\it instanton equations} gauge--fixing
the {\it topological symmetry} of the
topologically twisted dual N=2 theory \cite{topf4d_2,topf4d_3}.
They are produced in an algorithmic fashion
by applying the generalization derived in
\cite{topftwist_1,topftwist_2,topf4d_1} of
the twisting procedure of \cite{topfgen_7}.
\par
ii) Eq.s (\ref{mono}) are the
specialization to a very simple case, namely
to the case where both the world--manifold
${\cal M}$ and the target--manifold ${\cal N}$
are flat ${\IR}^4$, the gauge group $G$ is $U(1)$
and gravity is external,
of a more general set of three equations:
\begin{eqnarray}
R^{-ab}&=&-{1\over 2}I_u^{ab}q^*\hat\Omega^u,
\nonumber\\
F^{-ab}_\Lambda&=&-{g\over 2}I^{ab}_u
{\cal P}^u_\Lambda,\nonumber\\
{\cal D}_\mu q^i&-&{(j_u)_\mu}^\nu
{\cal D}_\nu q^j {(J_u)_j}^i=0,
\label{gin}
\end{eqnarray}
the first being the yield of dynamical gravity,
the last two being the appropriate generalization
of (\ref{mono}). Eq.s (\ref{gin}) are obtained by the twist
procedure of \cite{topftwist_1,topftwist_2,topf4d_1}
and are the gauged version of eq.s (2.1) and (3.12) of
\cite{topf4d_1}.
\par
iii) The interpretation of the third
of eq.s (\ref{gin}) as a Dirac  equation, as it
is done in the second of eq.s (\ref{mono}), is
a peculiarity of the flat case
${\cal N}=\IR^4$.
\par
The rest of this letter is devoted to prove the
above statements and to explain the symbols
appearing in eq.s (\ref{gin}).
\par
As just stated,
the puzzling feature of  conditions (\ref{mono})
is that scalar fields satisfy a  Dirac-type equation.
However, as it was shown in \cite{topf4d_1}, the second of
eq.s (\ref{mono}) has a  different geometrical meaning.
In the most general case, the scalars $M_\alpha$
(from now on to be denoted by $q^i$, $i=1,\ldots , 4$
in real notation)
describe a sigma model,
$q:{\cal M}\rightarrow {\cal N}$, mapping an almost
quaternionic four dimensional manifold
${\cal M}$\footnotemark\footnotetext{We recall
that this is not a restrictive requirement,
since any four dimensional
Riemannian manifold is almost quaternionic.}
into an almost quaternionic target manifold ${\cal N}$.
One can define a condition of {\sl triholomorphicity} for
the map $q$, that  arises naturally
from the topological twist of N=2 hypermultiplets
\cite{topf4d_1}.
For this reason, the solutions to that condition were named
by us hyperinstantons (= instantons of the hypermultiplets).
In the simplest case, namely ${\cal M}={\cal N}={\IR}^4$,
the instantonic equations reduce to the so called
Cauchy-Fueter equations \cite{hyperquotient_1}, that, after
introducing a quaternionic number notation, can be written in
a form that resembles the usual holomorphicity condition
\cite{topf4d_1}
\begin{equation}
\bar \partial q=0.
\label{fueter}
\end{equation}
It was well-known from the literature \cite{hyperquotient_1},
that these equations can also be written as
a Dirac equation on a spinor of definite chirality.
However, it was pointed out in \cite{topf4d_1} that this
is more a coincidence and does not correspond to the most
significative nature of the
equations\footnotemark\footnotetext{Actually, in the flat case
there are analogies with other equations of a known type.
For example, (\ref{fueter}) can be written as the self-duality
condition on the field strength of an abelian gauge-field in
the Lorentz  gauge.
Such misleading similarities explain why the generalization of
these equations given in \cite{topf4d_1} was not
straightforward.}.
What turned out to be the most significative interpretation
of the equations under consideration is that they are a
{\it triholomorphicity condition} on the map,
as already recalled.

In \cite{topf4d_1} the most general solution when
${\cal M}=T^4$ and ${\cal N}=T^{4\prime}$ are both four-tori
was found and the corresponding topological
$\sigma$-model was solved, leading to a meaningful partition function,
which turns out to be just a  $\theta$-function. This $\theta$-function is
characterized by a genus $g$, being integer valued between $0$ and
$12$. $g$ measures the {\sl degree of commensurability}
of the two tori. $g$ is a very nontrivial function on the $T^4$-moduli space,
and it would be very interesting to know it better.
In \cite{topf4d_1} examples with $g=0$, $g=12$ and $0<g<12$ where exhibited.

The reason why a Dirac--like equation makes its
appearance was also stressed
in \cite{topf4d_1}: it is the equation for the
infinitesimal deformations of
the triholomorphic map $q$ rather than the equation defining
the map $q$ itself.
This happens because the deformations of the map are the
topological ghosts, that, {\sl via} topological twist,
come from the fermions of the hypermultiplet
(the {\sl hyperini}). Of course, the field equations of such
fermions are Dirac-type equations. With a flat target manifold,
the triholomorphicity condition is linear in the map $q$,
so that the equations of the deformations of the map have the
same  form as the equations of the map itself, and that is why
they resemble the Dirac equation.
Thus, the second of eq.s (\ref{mono}) looks like the gauged
version of the hyperinstanton equations. Therefore,
the solutions to this type of equations
will be from now on  named {\sl gauged hyperinstantons}.

It remains to explain the first of eq.s (\ref{mono}),
that looks more mysterious.
Actually, it is much less mysterious if we recall one more result
obtained in \cite{topf4d_1}.
There it was shown that  when hyperinstantons are
coupled to dynamical gravity, they modify the equation
$R^{-ab}=0$ of gravitational instantons as follows
\begin{equation}
R^{-ab}=-{1\over 2}I_u^{ab}q^*\Omega^u,
\label{eq}
\end{equation}
which is the generalization of the self-duality condition on
the Riemann tensor.
Here $I^u_{ab}$, $u=1,2,3$ is a triplet of
$4\times 4$ antiselfdual matrices ($I^u_{ab}=I^{u-}_{ab}$)
satisfying the quaternionic algebra
(see formula (\ref{complex})). $\Omega^u$ will be defined
below. If the target manifold is four dimensional, one can
also write $R^{-ab}=\tilde R^{-ab}$, where the tilded
$2$--form is the pull-back  of the corresponding target
$2$--form\footnotemark\footnotetext{In Minkowskian notation, we use
$M^{-ab}\equiv{1\over 2}\left(M^{ab}+{i\over
2}\varepsilon^{abcd}M^{cd}\right)$, while in Euclidean notation we use
$M^{-ab}\equiv{1\over 2}\left(M^{ab}-{1\over
2}\varepsilon^{abcd}M^{cd}\right)$.}.

This remark suggests that when hyperinstantons
are coupled to Yang-Mills fields, instead of gravity,
the Yang-Mills instanton condition $F^-_{\mu\nu}=0$ should be
modified in a similar way.
Thus, the first of eq.s (\ref{mono}) is the gauge-analogue of
the modification (\ref{eq}) that is present is the gravitational
case. It is clear from the argument developed so far
that the most general form of the first of conditions
(\ref{mono}) can be found by repeating the twisting
exercise of ref. \cite{topf4d_1} in the gauged
case\footnotemark\footnotetext{We recall that
the twisting procedure of ref.\ \cite{topf4d_1},
which was firstly defined in \cite{topftwist_2},
is a nontrivial generalization of that of
\cite{topfgen_7}, since the procedure  of \cite{topfgen_7}
could not work on hypermultiplets.
In particular, in \cite{topf4d_1} it was shown that one also has
to identify $SU(2)_L$ with a suitable $SU(2)$ subgroup of the Lorentz
group of the target manifold ${\cal N}$.}.
The rest of this letter is devoted to derive the
most general form (\ref{gin}) of the gauged
hyperinstanton equations and to show that they reduce to
(\ref{mono}) in the simplest case.
The interpretation offered here also provides
expressions of the topological  observables of the theory,
since the topological field theory  encoded in (\ref{mono}) is
nothing else but a particular case of the known topological
models  (see \cite{topf4d_1} for more details).
\par
By definition, ${\cal N}$ possesses an almost quaternionic
structure, namely three locally
defined\footnotemark\footnotetext{It means that
the almost quaternionic $(1,1)$-tensors are defined on
neighborhoods $U_{(\alpha)}$
such that on the intersection $U_{(\alpha)}\cap U_{(\beta)}$
of two neighborhoods the transition functions
are $SO(3)$ matrices $\Lambda^{uv}$ \cite{momentmap_1}.}
$(1,1)$-tensors $J^u$, $u=1,2,3$, satisfying the quaternionic
algebra
\begin{equation}
J^u J^v= - \delta^{uv}+\varepsilon^{uvz}J^z.
\label{complex}
\end{equation}
Moreover, ${\cal N}$ is endowed with a metric
$h_{ij}$ that is by assumption  Hermitean with respect to the
almost quaternionic $(1,1)$-tensors $J^u$.
One can introduce the generalized K\"ahler forms
\begin{equation}
\Omega^u=\lambda h_{ik}{(J^u)_j}^kdq^i\wedge dq^j.
\label{2.56}
\end{equation}
In particular, if the $J_u$ are  globally defined and covariantly
constant complex structures then the target manifold ${\cal N}$
is a hyperK\"ahler manifold.
In that case, the forms $\Omega^u$ are closed:
\begin{equation}
d\Omega^u=0.
\label{7}
\end{equation}
On the other hand, if ${\cal N}$ is a quaternionic
K\"ahler manifold,  there exist three one-forms $\omega^u$ that
make an $SU(2)$ connection, with respect to which
the forms $\Omega^u$ are covariantly closed and such that
$\Omega^u$ is the field strength of this connection.
To say it in formulae, we have:
\begin{equation}
d\Omega^u+\varepsilon^{uvz}\omega^v\wedge\Omega^z=0,
\quad\quad
d\omega^u+{1\over 2}\varepsilon^{uvz}
\omega^v\wedge\omega^z=\Omega^u.
\label{8}
\end{equation}
Any quaternionic K\"ahler manifold is an Einstein manifold.
Then, in (\ref{2.56}) $\lambda$ is a real constant that
is related to  the cosmological constant of ${\cal N}$.
When the limit  $\lambda\rightarrow 0$ is taken in an
appropriate way \cite{topf4d_1}, then one can go from the
quaternionic K\"ahler to the hyperK\"ahler case.

As far as ${\cal M}$ is concerned, since it is four
dimensional, it is sufficient to have a metric $g_{\mu\nu}$
to endow it with an almost quaternionic structure, as we
already recalled. Explicitly, we have
\begin{equation}
(j_u)_\mu^\nu=I_u^{ab}\, e_{a\mu} \, e_b^{\nu},
\end{equation}
$e^a_\mu$ being the vierbein.
The instantons as derived from the topological twist of
\cite{topftwist_2,topf4d_1} are given by the following
condition on the world metric:
\begin{equation}
{\omega^-}^{ab}=-{1\over 2}I_u^{ab}q^*\omega^u,
\label{inst2}
\end{equation}
that is equivalent to (\ref{eq}) in a suitable local Lorentz
frame\footnotemark\footnotetext{(\ref{inst2}) is the form of
(\ref{eq}) as obtained by the twist.  It is well-defined
only if ${\cal N}$ is  quaternionic K\"ahler (or
hyperK\"ahler, in which case the right hand side of
(\ref{inst2}) is zero). On the other hand, (\ref{eq})
is well-defined in the most general case.},
plus the following equations on the map
$q:{\cal M}\rightarrow {\cal N}$,
\begin{equation}
e^{\mu[a}E_i^{b]^+k}\partial_\mu q^i =0,\quad\quad
e^\mu_a E^{ak}_i \partial_\mu q^i =0,
\label{inst1}
\end{equation}
$[ab]^+$ meaning antisymmetrization and self--dualization
in the indices $a,b$\footnotemark\footnotetext{Notice
that when in (\ref{inst2}) the duality is $-$,
then in (\ref{inst1})
it is necessarily $+$. This is a consequence of a $U(1)$
symmetry discovered in \cite{topftwist_2}. It makes the improved
topological twist meaningful, since it defines the new ghost number.
It is called R-duality and generalizes the R-symmetry to supergravity.}.
The vielbein $E_i^{ak}$ of the target
manifold ${\cal N}$, defined so that $h_{ij}=2E^{ak}_iE^{ak}_j$,
has a Lorentz index that is split into $(a,k)$,
$a$ being identified with the Lorentz index of the world
manifold and
$k$ being an extra index ranging from $1$ to $n$,
if ${\rm dim}\, {\cal N}=4n$.
This is the effect of the topological twist of
\cite{topf4d_1}.
As a matter of fact, written in this form,
(\ref{inst1}) are not sufficiently explicit. Introducing
the inverse vielbein $E^i_{ak}$ ($E^i_{ak}E^{ak}_j=\delta^i_j$,
$E^i_{ak}E^{bl}_i=\delta^b_a\delta^l_k$)
and the almost quaternionic $(1,1)$ tensors
\begin{equation}
{(J_u)_i}^j=(I_u)_a^bE^{ak}_iE^j_{bk},
\end{equation}
(\ref{inst1}) becomes (see \cite{topf4d_1} for the details):
\begin{equation}
\partial_\mu q^i-{(j_u)_\mu}^\nu\partial_\nu q^j {(J_u)_j}^i=0,
\label{afeq2}
\end{equation}
which appears clearly as a generalization of the Cauchy-Riemann
equations\footnotemark\footnotetext{The number of independent conditions
contained
in (\ref{afeq2}) is equal to ${\rm dim}\, {\cal N}$, as it must be. This
follows from
a duality condition satisfied identically by the matrix
$H_\mu^i\equiv\partial_\mu q^i-{(j_u)_\mu}^\nu\partial_\nu q^j {(J_u)_j}^i$,
namely
$H_\mu^i+{1\over 3}{(j_u)_\mu}^\nu H_\nu^j {(J_u)_j}^i=0$ \cite{topf4d_1}.}.
These equations are a condition of triholomorphicity
of the maps ${\cal M}\rightarrow {\cal N}$ and that
is why we named
triholomorphic a map $q$ satisfying eq.s (\ref{afeq2}).

As a matter of fact, the contraction between the indices $u$
of the almost quaternionic structures on the  two manifolds
can be performed  introducing an arbitrary point--dependent
$SO(3)$ matrix $\Lambda^{uv}$, since an almost quaternionic structure is
defined up to $SO(3)$ matrices:
\begin{equation}
\partial_\mu q^i-\Lambda^{uv}{(j_u)_\mu}^\nu\partial_\nu q^j
{(J_v)_j}^i=0.
\label{lambda}
\end{equation}
The solutions can be called $\Lambda$-triholomorphic.
Some properties of this ambiguity have been also studied
in \cite{topf4d_1} on explicit
examples of isometries for ${\cal M}={\cal N}=K3$
(to be precise in the realization of K3 as a Fermat
surface in $\IC\IP_3$ [4]).

It is clear that when identifying indices
of the Lorentz groups of two different manifolds, one has to be careful
about covariance. The role of (\ref{inst2}) is then to
relate the spin connections of the two manifolds consistently:
one can no more distinguish an index $u$ for ${\cal M}$ and
one for ${\cal N}$; similarly, the corresponding components of the
spin connections of ${\cal M}$ and ${\cal N}$ are identified, so that
it is immaterial which one is used in defining the covariant
derivative for $u$-indexed tensors. Stated in a different way,
(\ref{inst2}) is the condition for making $I^u_{ab}$ covariantly constant:
\begin{equation}
{\cal D}(I^u)^{ab}=d(I^u)^{ab}-\omega^{-ac}(I^u)^{cb}+
\omega^{-bc}(I^u)^{ca}+\varepsilon^{uvz}q^*\omega^v (I^z)^{ab}=0.
\end{equation}
\par
The purpose, now, is to gauge the hyperinstanton equations.
So, suppose that the target manifold ${\cal N}$ admits
Killing vectors
\begin{equation}
k_\Lambda(q)=k_\Lambda^i(q){\partial\over \partial q^i},
\quad\quad\quad
[k_\Lambda,k_\Sigma]=- f^\Gamma_{\Lambda\Sigma}k_\Gamma,
\label{kvectors}
\end{equation}
of a certain Lie algebra ${\cal G}$ with structure constants
$f^\Gamma_{\Lambda\Sigma}$.
Then, one can introduce the covariant derivatives
\begin{equation}
{\cal D}q^i=dq^i+g A^\Lambda k^i_\Lambda(q)
\end{equation}
and replace (\ref{inst1})  with
\begin{equation}
e^{\mu[a}E_i^{b]^+k} {\cal D}_\mu q^i =0,\quad\quad
e^\mu_a E^{ak}_i {\cal D}_\mu q^i =0,
\label{ginst1}
\end{equation}
or, equivalently, (\ref{afeq2}) with
\begin{equation}
{\cal D}_\mu q^i-{(j_u)_\mu}^\nu {\cal D}_\nu q^j {(J_u)_j}^i=0.
\label{gafeq2}
\end{equation}
(\ref{ginst1}) is in agreement with the topological twist
of \cite{topftwist_2,topf4d_1}
when applied to the gauged N=2 supersymmetric $\sigma$-model
\cite{skgsugra_1}. The solutions $q$ to (\ref{gafeq2}) can be called {\sl
gauged triholomorphic}
maps.
The same twisting procedure provides the
generalization of the first equation of (\ref{mono}).
Here, we shall consider the most general case,
in which gravity is dynamical, so that we shall also
find the (straightforward) generalization of (\ref{eq}).
Following \cite{skgsugra_1}, we see that the gauging
is achieved with the replacements
\begin{eqnarray}
\omega^u&\rightarrow &\hat\omega^u=\omega^u+g A^\Lambda{\cal
P}^u_\Lambda,\nonumber\\
\Omega^u&\rightarrow & \hat\Omega^u=d\hat\omega^u+{1\over
2}\varepsilon^{uvz}\hat\omega^v\hat \omega^z=
\Omega^u_{ij}{\cal D}q^i\wedge
{\cal
D}q^j+gF^\Lambda{\cal P}^u_\Lambda,
\label{18}
\end{eqnarray}
 where ${\cal P}^u_\Lambda$ is the {\it momentum map}
function,
while $F^\Lambda$ is the field strength of
 the gauge-field,
\begin{equation}
F^\Lambda=dA^\Lambda+{1\over 2}
g{f^\Lambda}_{\Sigma\Gamma}A^\Sigma
A^\Gamma.
\end{equation}
Let us then pause for a moment and recall the
important notion of momentum map \cite{momentmap_1}.
\par
\underline{\sl Momentum map for hyperK\"ahler manifolds}
\par\noindent
Consider a compact Lie group $G$ acting on a hyperK\"ahler
manifold ${\cal N}$ of
real dimension $4n$ by means of Killing vector fields ${\bf X}$ that are
holomorphic
with respect to the three complex structures of ${\cal N}$;
then these vector
fields preserve also the K\"ahler forms:
\begin{equation}
\left.\begin{array}{l}
{\cal L}_{\scriptscriptstyle{\bf X}}g = 0 \leftrightarrow
\nabla_{(\mu}X_{\nu)}=0 \\
{\cal L}_{\scriptscriptstyle{\bf X}}J^u = 0 \,\, , \,u =1,2,3\\
\end{array}\right\} \,\,\Rightarrow\,\,
0={\cal L}_{\scriptscriptstyle{\bf X}}
\Omega^u = i_{\scriptscriptstyle{\bf X}}
d\Omega^u+d(i_{\scriptscriptstyle{\bf X}}
\Omega^u) = d(i_{\scriptscriptstyle{\bf X}}\Omega^u)\, .
\label{holkillingvectors}
\end{equation}
Here ${\cal L}_{\scriptscriptstyle{\bf X}}$ and
$i_{\scriptscriptstyle{\bf X}}$ denote respectively
the Lie derivative along the vector field ${\bf X}$ and the
 contraction (of forms) with it.
If ${\cal N}$ is simply connected,
$d(i_{{\bf X}}\Omega^u)=0$ implies the global existence
of three functions ${\cal P}^u_{{\bf X}}$ such that
\begin{equation}
i_{\scriptscriptstyle{\bf X}}\Omega^u=
-d{\cal P}^{u}_{{\bf X}}.
\label{amomentimappo1}
\end{equation}
If ${\cal N}$ is not simply connected, the functions ${\cal P}^u_{\bf X}$ exist
only locally.
The  ${\cal P}^{u}_{{\bf X}}$ are
defined up to a constant,
which can be arranged so as to make them equivariant
\begin{equation}
{\bf X} {\cal P}^{u}_{\bf Y} =
2\,\Omega^{u} ({\bf X} , {\bf Y} ) =
{\cal P}^{u}_{[{\bf X},{\bf Y}]}.
\label{equivariance}
\end{equation}
The $\{{\cal P}^{u}_{{\bf X}}\}$ constitute then a
{\it momentum map}.
This can be regarded as a map
${\cal P}: {\cal N} \Rightarrow {\IR}^3\otimes
{\cal G}^*$, where ${\cal G}^*$ denotes the dual
 of the Lie algebra ${\cal G}$ of the
group $G$. Indeed let $x\in {\cal G}$ be the
 Lie algebra element corresponding to the Killing
vector ${\bf X}$; then, for a given $m\in {\cal N}$,
the functional ${\cal P}^{u}(m) : x \longrightarrow
{\cal P}^{u}_{{\bf X}}(m) \in \IC$ is a linear
functional on ${\cal G}$. In practice,
expanding ${\bf X} =X^\Lambda {\bf k}_\Lambda$
on a basis of Killing vectors ${\bf k}_\Lambda$
such that (\ref{kvectors}) holds,
we also have  ${\cal P}^{u}_{{\bf X}}=X^\Lambda \,
{\cal P}^{u}_\Lambda$, $u=1,2,3$;
the ${\cal P}^{u}_\Lambda$ are the components of
the momentum map.
\par
\underline{\sl Momentum map for quaternionic K\"ahler manifolds}
\par\noindent
In the case of a quaternionic K\"ahler manifold
where the three $2$--forms
$\Omega^{u}$ are not closed but just covariantly
closed with respect to
the $SU(2)$ connection $\omega^{u}$, then the momentum
map also exists but equation (\ref{amomentimappo1})
 is replaced by its
$SU(2)$--covariant analogue:
\begin{equation}
i_{\scriptscriptstyle{\bf X}}\Omega^u=
-\nabla{\cal P}^{u}_{{\bf X}}=
-\left ( d{\cal P}^{u}_{{\bf X}}\, + \varepsilon^{uvz}
\omega^{v} {\cal P}^{z}_{{\bf X}}
\right ).
\label{amomentimappo2}
\end{equation}
Now ${\cal P}^u_{\bf X}$ are fixed uniquely \cite{hyperquotient_1}.
They satisfy identically the following
modified equivariance condition
\begin{equation}
{\cal P}^{u}_{[{\bf X},{\bf Y}]}-
2\, \Omega^{u}({\bf X} , {\bf Y})
+\varepsilon^{uvz}{\cal P}^{v}_{\bf X}
 {\cal P}^{z}_{\bf Y}=0,
\label{covequivariance}
\end{equation}
that generalizes  (\ref{equivariance}). This is proven  by showing that,
calling $C^u_{\Lambda\Sigma}$ the left hand side of (\ref{covequivariance}),
one has $\nabla C^u_{\Lambda\Sigma}=0$ \cite{skgsugra_1}. Then, the result
follows from
$0=\nabla^2 C^u_{\Lambda\Sigma}=\varepsilon^{uvz}\Omega^v C^z_{\Lambda\Sigma}$.
\par
\vspace{5pt} \par
Equipped with these results (for more
details see \cite{momentmap_1} and \cite{skgsugra_1})
we can now resume our previous discussion.
(\ref{covequivariance}) reads
\begin{equation}
0=\Omega^u_{ij}k^i_\Sigma k^j_\Gamma+
{1\over 2}{f^\Lambda}_{\Sigma\Gamma}
{\cal P}^u_\Lambda-{1\over 2}
\varepsilon^{uvz}{\cal P}^v_\Sigma{\cal P}^z_\Gamma
\label{gcond}
\end{equation}
and guarantees the consistency of (\ref{18})
and (\ref{amomentimappo2}) \cite{skgsugra_1}.
The relevant supersymmetry transformation
 is the one of the right handed gaugino \cite{skgsugra_1}
\begin{equation}
s\lambda^I_A=\cdots+{1\over 2}F^{I-}_{ab}
\gamma^{ab}\varepsilon_{AB}C^B
+ig{(\sigma_u)_A}^C\varepsilon_{BC}C^B{\cal P}^{Iu}.
\end{equation}
In this equation, the scalars of the vector multiplets
have been set to zero,
since they become ghosts for the ghosts after the twist.
Similarly, the graviphoton
has to be set to zero \cite{topftwist_1}.
 For this reason, the index $I$ is the same
as the index $\Lambda$. $C^B$ are the right handed
components of the supersymmetry ghosts.
Performing the topological twist formulated in
\cite{topftwist_2}, one finds the
following instantonic condition:
\begin{equation}
F^{-ab}_\Lambda=-{g\over 2}I^{ab}_u{\cal P}^u_\Lambda,
\label{29}
\end{equation}
which is the desired generalization of
the first of (\ref{mono}).
The generalization of (\ref{eq}), on the other hand,
is obtained by replacing
$\Omega^u$ with $\hat \Omega^u$. Summarizing,
the complete set of
equations for the gauged hyperinstantons are given by
eq.s (\ref{gin}),
the first of which can be also expressed,
when ${\cal N}$ is quaternionic K\"ahler, in the form
\begin{equation}
{\omega^-}^{ab}=-{1\over 2}I_u^{ab}q^*\hat\omega^u.
\label{gin2}
\end{equation}
The total  ghostless  twisted lagrangian of the most
general N=2 theory\footnotemark\footnotetext{This the
bosonic lagrangian with the graviphoton and the scalar fields
of the vector multiplets equated to zero, since
after the twist they become ghost fields.} is:
\begin{eqnarray}
{\cal L}&=&\varepsilon_{abcd}R^{ab}e^c e^d
-{1\over 6}\lambda g^{\mu\nu}h_{ij}{\cal D}_\mu
q^i{\cal D}_\nu q^j
\, \varepsilon_{abcd}e^a e^b e^c e^d
\nonumber\\
&-&{1\over 12}(F^{ab}_\Lambda F_\Lambda^{ab}
+2g^2 {\cal P}^u_\Lambda {\cal P}^u_\Lambda)
\, \varepsilon_{abcd}e^a e^b e^c e^d,
\end{eqnarray}
the last term being the scalar potential. A crucial test for conditions
(\ref{gafeq2}), (\ref{29}) and (\ref{gin2}) is to show that
${\cal L}$ can be written as the sum of their squares
plus a topological term and a total derivative, namely
\begin{eqnarray}
{\cal L}&=&4i\left(\omega^{-ab}+{1\over 2}I^{ab}_u q^*\hat\omega^u\right)\wedge
\left(\omega_{-ac}+{1\over 2}({I_v})_{ac} q^*\hat\omega^v\right)
e^b e^c
\nonumber\\
&-&{\lambda\over 24}\,
g^{\mu\nu}h_{ij}
\left({\cal D}_\mu q^i-{(j_u)_\mu}^\rho {\cal D}_\rho q^k
{(J_u)_k}^i\right)
\left({\cal D}_\nu q^j-{(j_v)_\nu}^\sigma {\cal D}_\sigma q^l
{(J_v)_l}^j\right)\,
\varepsilon_{cdef}e^ce^de^ee^f\nonumber\\
&-&{1\over 6}\left(F^{-ab}_\Lambda+{g\over 2}I^{ab}_u{\cal P}^u_\Lambda
\right)^2\, \varepsilon_{cdef}e^c e^d e^e e^f
\nonumber\\&-&
iF_\Lambda F_\Lambda-4i  d\left[\left(\omega^{-ab}+{1\over 2}
I_u^{ab}q^*\hat\omega^u\right)e^a e^b\right].
\label{32}
\end{eqnarray}
This proves that the solutions to (\ref{gin}) are solutions to the
Einstein-Yang-Mills-matter-coupled field equations.
Notice that the last total derivative term of (\ref{32}) is zero for
any hyperinstanton. On the solutions of (\ref{gin}),(\ref{gin2})
the action $S$ is simply
\begin{equation}
S=-i\int_{\cal M}{\rm tr}\,[F\wedge F],
\end{equation}
i.e.\  the Pontrjiagin number of the gauge bundle.

The observables encoding the meaningful topological invariants
will be not written down explicitly, due to lack of space.
They are the observables of the $\sigma$-model \cite{topf4d_1},
topological gravity \cite{topftwist_1}
and topological Yang-Mills theory \cite{topfgen_7,me},
coupled together (see \cite{me} for explicit examples
of nontrivial couplings).

One can formally turn to the case when supersymmetry is global
(${\cal N}$ hyperK\"ahler)
by performing the following replacements \cite{topf4d_1}:
\begin{equation}
A^\Lambda \rightarrow   \lambda^{1\over2}A^\Lambda,\quad
g\rightarrow \lambda^{-{1\over 2}}g,\quad
\Omega^u\rightarrow \lambda \Omega^u,\quad
\omega^u\rightarrow {\cal O}(\lambda),\quad {\cal P}^u_\Lambda\rightarrow
\lambda
{\cal P}^u_\Lambda,
\end{equation}
and simplifying $\lambda$ wherever possible. At the
end one puts $\lambda=0$. In this way (\ref{8}),
(\ref{amomentimappo2})
and (\ref{covequivariance}) become (\ref{7}),
(\ref{amomentimappo1}) and (\ref{equivariance}),
respectively. The first of (\ref{gin})
becomes $R^{-ab}=0$, so that ${\cal M}$
is also hyperK\"ahler. Finally, $\hat \Omega^u$ become closed,
$d\hat \Omega^u=0$. The lagrangian
\begin{equation}
{\cal L}=
-{1\over 12} [2g^{\mu\nu}h_{ij}{\cal D}_\mu q^i{\cal D}_\nu q^j+
F^{ab}_\Lambda F_\Lambda^{ab}+2g^2 {\cal P}^u_\Lambda
{\cal P}^u_\Lambda]\, \varepsilon_{cdef}e^c e^d e^e e^f
\label{LK}
\end{equation}
can then be written as
\begin{eqnarray}
{\cal L}&=&-{1\over 24}
g^{\mu\nu}h_{ij}
\left({\cal D}_\mu q^i-{(j_u)_\mu}^\rho {\cal D}_\rho q^k
{(J_u)_k}^i\right)
\left({\cal D}_\nu q^j-{(j_v)_\nu}^\sigma {\cal D}_\sigma q^l
{(J_v)_l}^j\right)\,
\varepsilon_{cdef}e^ce^de^ee^f\nonumber\\
&-&{1\over 6}\left(F^{-ab}_\Lambda+{g\over 2}I^{ab}_u{\cal P}^u_\Lambda
\right)^2\, \varepsilon_{cdef}e^c e^d e^e e^f
-iF_\Lambda F_\Lambda-2i\Theta^u \hat \Omega^u,
\label{action}
\end{eqnarray}
where $\Theta^u=I^u_{ab}e^ae^b$ are the K\"ahler forms of ${\cal M}$.
The last term of (\ref{action}) is also a
topological invariant, like in \cite{topf4d_1},
since both $\Theta^u$ and $\hat \Omega^u$ are closed.
Consequently, (\ref{LK}) is minimized by the second and the third conditions of
(\ref{gin}):
\begin{equation}
F_\Lambda^{-ab}+{g\over 2}I^{ab}_u{\cal P}^u_\Lambda=0,\quad\quad
{\cal D}_\mu q^i-{(j_u)_\mu}^\nu {\cal D}_\nu q^j {(J_u)_j}^i=0.
\label{ki}
\end{equation}
{}From these equations, we can now  retrieve eq.s (\ref{mono}) explicitly,
choosing ${\cal M}={\cal N}=\IR^4$ and $G=U(1)$.
Let us consider a Killing vector of the form
\begin{equation}
k(q)=M_{ij}q^i{\partial\over \partial q^j}=k^i{\partial\over \partial q^i},
\end{equation}
$M$ being a to-be-determined $4\times 4$ constant matrix.
We have $\Omega^u= I^u_{ij}dq^i\wedge dq^j$. In Euclidean notation,
we choose the matrices $I^u_{ab}$ as
follows \cite{topf4d_1}:
\begin{equation}
\matrix{I_1=\left(\matrix{0&1&0&0\cr -1&0&0&0\cr
0&0&0&-1\cr 0&0&1&0}\right),&
I_2=\left(\matrix{0&0&1&0\cr 0&0&0&1\cr -1&0&0&0\cr
0&-1&0&0}\right),&
I_3=\left(\matrix{0&0&0&1\cr 0&0&-1&0\cr  0&1&0&0\cr
-1&0&0&0}\right).}
\label{opuwe}
\end{equation}
Equation (\ref{amomentimappo1}) gives
\begin{equation}
{\partial {\cal P}^u\over \partial q^i}=-2 q^kM^{kj}I^u_{ji}.
\end{equation}
The integrability condition, i.e.\ (\ref{holkillingvectors}),
requires $MI^u$ to be symmetric $\forall u$.
A good $M$ is any matrix with the opposite duality of $I^u$, for example,
\begin{equation}
\bar I=\left(\matrix{0&1&0&0\cr -1&0&0&0\cr
0&0&0&1\cr 0&0&-1&0}\right),\quad\quad \bar I=\bar I^+.
\end{equation}
Then one finds
\begin{equation}
{\cal P}^u=-q^t\bar I I^uq,\quad\quad\quad k=q^t\bar I{\partial\over \partial
q},
\end{equation}
$t$ meaning transposition.
The equivariance condition (\ref{equivariance}) is trivially satisfied.
With the same identification as in \cite{topf4d_1}, namely
\begin{equation}
\psi=\left[\matrix{0\cr 0\cr q^4-iq^3\cr q^2-iq^1}\right]=
\left[\matrix{0\cr 0\cr M^1\cr M^2}\right],\quad\quad
\gamma_i=\left(\matrix{0&i\sigma_i\cr -i\sigma_i&0}\right),\quad
\gamma_4=\left(\matrix{0&1\cr 1&0}\right),
\end{equation}
the second equation of (\ref{ki}) becomes the Dirac equation
\begin{equation}
{\cal D}\!\!\!\!\slash \psi=\gamma^\mu{\cal
D}_\mu\psi=\gamma^\mu(\partial_\mu+igA_\mu)\psi=0.
\end{equation}
Instead, the first of (\ref{ki}) becomes
\begin{eqnarray}
F^-_{12}&=&{g\over 2}[(q^4)^2+(q^3)^2-(q^2)^2-(q^1)^2]={g\over 2}(M^1\bar
M^1-M^2\bar
M^2),\nonumber\\
F^-_{13}&=&g(q^1q^4-q^2q^3)=i{g\over 2}(M^2\bar M^1-M^1\bar M^2),\nonumber\\
F^-_{23}&=&g(q^1q^3+q^2q^4)={g\over 2}(M^2\bar M^1+M^1\bar M^2),
\end{eqnarray}
which is equivalent to the first of (\ref{mono}).
Finally, with a triplet $\bar I_u=\bar I_u^{+}$ satisfying (\ref{complex}) and
${\cal P}^u_v=-q^t\bar I_v I^u q$, one can consider the case
${\cal M}={\cal N}=\IR^4$, $G=SU(2)$. Again, (\ref{equivariance}) is easily
checked.


\end{document}